\documentclass[prl,aps,showpacs,twocolumn,superscriptaddress,longbibliography]{revtex4-2}
\usepackage{xcolor}
\usepackage{bm}
\usepackage{amsmath}
\usepackage{amsthm}
\usepackage{stmaryrd}
\usepackage{graphicx}
\usepackage{physics}
\usepackage{bbding}
\usepackage{dsfont}
\usepackage[colorlinks=true,linkcolor=blue,anchorcolor=blue,citecolor=blue,urlcolor=blue]{hyperref}

\begin{document}

\title{Quasisymmetry Enriched Gapless Criticality at Chern Insulator Transitions}

\author{Jiayu Li}
\affiliation{New Cornerstone Science Lab, Department of Physics, The University of Hong Kong, Hong Kong, China}
\affiliation{HK Institute of Quantum Science \& Technology, The University of Hong Kong, Hong Kong, China}
\affiliation{State Key Laboratory of Optical Quantum Materials, The University of Hong Kong, Hong Kong, China}

\author{Feng-Ren Fan}
\affiliation{School of Physical Science and Technology, Soochow University, Suzhou 215006, China}

\author{Wang Yao}
\email{wangyao@hku.hk}
\affiliation{New Cornerstone Science Lab, Department of Physics, The University of Hong Kong, Hong Kong, China}
\affiliation{HK Institute of Quantum Science \& Technology, The University of Hong Kong, Hong Kong, China}
\affiliation{State Key Laboratory of Optical Quantum Materials, The University of Hong Kong, Hong Kong, China}

\date{\today}

\begin{abstract}
In continuous topological phase transitions (CTPTs), the low-energy physics is governed by gap-closing subspaces, where approximate ``higher" symmetries, termed quasisymmetries, may emerge. Here, we introduce the notion of quasisymmetry enrichment of these transitions. Focusing on paradigmatic normal-to-Chern insulator transitions, we identify quasisymmetries in the gapless subspaces, which subdivide CTPTs of the same universality class according to quasisymmetry charges. Gapless criticalities with nontrivial charges exhibit regulated phenomena, including intrinsic correlations between charge and pseudospin currents and continuous generalized Hall conductivities governed by the generalized Středa formula, both conventionally exclusive to gapped phases. These features arise as  quasisymmetry forbids certain matrix elements, rendering the generalized Berry curvature integrable. By establishing quasisymmetry as a fundamental classifying ingredient, our work adds a new dimension for understanding the rich landscape of quantum phase transitions. 
\end{abstract}

\maketitle

Topological phase transitions (TPTs) signify fundamental changes in the topological properties of quantum states of matter. In gapped systems such as insulators, distinct phases on either side of a TPT are classified by robust bulk topological invariants ~\cite{SchnyderPRB2008Classificationtopologicalinsulators,Kitaev2009Periodictabletopological,HasanRoMP2010ColloquiumTopologicalinsulators,RyuNJoP2010Topologicalinsulatorssuperconductors,QiRoMP2011Topologicalinsulatorssuperconductors,AndoARoCMP2015TopologicalCrystallineInsulators,KruthoffPRX2017TopologicalClassificationCrystalline,BradlynN2017Topologicalquantumchemistry,TangN2019Comprehensivesearchtopological,VergnioryN2019completecataloguehigh,ZhangN2019Cataloguetopologicalelectronic,ElcoroNC2021Magnetictopologicalquantum}.
In parallel, it is equally essential to classify their criticality—the universal properties of the quantum critical point itself—to fully unravel the nature of these transitions.
Conventionally, this classification is approached from two perspectives: topologically, by discrete jumps of topological invariants such as the Chern number in Chern insulator transitions~\cite{ThoulessPRL1982QuantizedHallConductance,HaldanePRL1988ModelQuantumHall}; and energetically, by the occurrence of  energy gap closure, distinguishing between first-order and continuous transitions \cite{RoyPRB2016Continuousdiscontinuoustopological,WenRoMP2017ColloquiumZoo}. In weak-interacting systems, TPTs are continuous with energy gap closure, which can be further subdivided as distinct universality classes depending on the spatial dimension and specific band crossing types (e.g., nodal point or line) \cite{ChenPRB2017Correlationlengthuniversality}. Strong interactions, on the other hand, can induce first order TPTs without gap closure \cite{AmaricciPRL2015FirstOrderCharacter,ImriskaPRB2016Firstordertopological,RoyPRB2016Continuousdiscontinuoustopological,BarbarinoPRB2019Firstordertopological}.

Beyond these established paradigms, the role of symmetry protection or symmetry enrichment has emerged as a crucial new dimension for classifying the gapless criticalities of continuous TPTs (CTPTs)~\cite{ScaffidiPRX2017GaplessSymmetryProtected, VerresenPRX2021GaplessTopologicalPhases,YuPRL2022ConformalBoundaryConditions,ChatterjeePRB2023Holographictheorycontinuous,WenPRB2023Bulkboundarycorrespondence}. In this framework, gapless criticalities within the same universality class can further split into distinct subclasses labeled by symmetry charges~\cite{VerresenPRX2021GaplessTopologicalPhases}. Remarkably, those with nontrivial symmetry charges can host phenomena typically associated with gapped phases only, e.g. protected boundary modes, now precisely at the gapless quantum critical point~\cite{VerresenPRX2021GaplessTopologicalPhases,YuPRL2022ConformalBoundaryConditions}.

Recently, the concept of quasisymmetry has been introduced to describe additional symmetries that act exclusively within a relevant subspace of the full Hamiltonian's Hilbert space~\cite{GuoNP2022Quasisymmetryprotected,LiPRL2024GroupTheoryQuasisymmetry}. Although not exact symmetries of the Hamiltonian, quasisymmetries can protect near-degenerate bands \cite{GuoNP2022Quasisymmetryprotected,LiPRL2024GroupTheoryQuasisymmetry}, quantum many-body scars \cite{RenPRL2021QuasisymmetryGroupsMany}, new quasiparticle \cite{ZhangN2025Topologicalchargequadrupole}, near-quantized spin Hall effect \cite{LiuPRB2024QuantumspinHall}, and weak ferromagnetism in altermagnets \cite{RoigPRL2025QuasisymmetryConstrainedSpin}. This raises a natural question: in CTPTs, where low-energy physics at criticality is governed by the gapless subspace that may host quasisymmetries, can quasisymmetries likewise enrich the classification of the transition?

In this Letter, we answer this question affirmatively. Using Chern insulator (CI) transitions as examples, we demonstrate that quasisymmetries emerging within the gap-closing subspace can indeed enrich the transitions by endowing the gapless criticality with a quasisymmetry charge. Based on the charge values, the CTPTs within the same universality class are sub-classified. CTPTs with trivial charge behave as if quasisymmetry were absent. 
In contrast, criticalities with nontrivial charge exhibit protected gapped phenomena, manifested through an intrinsic correlation between charge and pseudospin currents, and a continuous generalized Hall conductivity associated with a generalized Středa formula. Such phenomena—conventionally hallmarks of insulating phases—persist at the gapless critical point due to quasisymmetry-forbidding certain matrix element in the generalized Berry curvature. Our results establish quasisymmetry as a fundamental new ingredient for classifying and understanding the rich landscape of CTPTs.

\begin{figure}
\centering
\includegraphics[width=1\columnwidth]{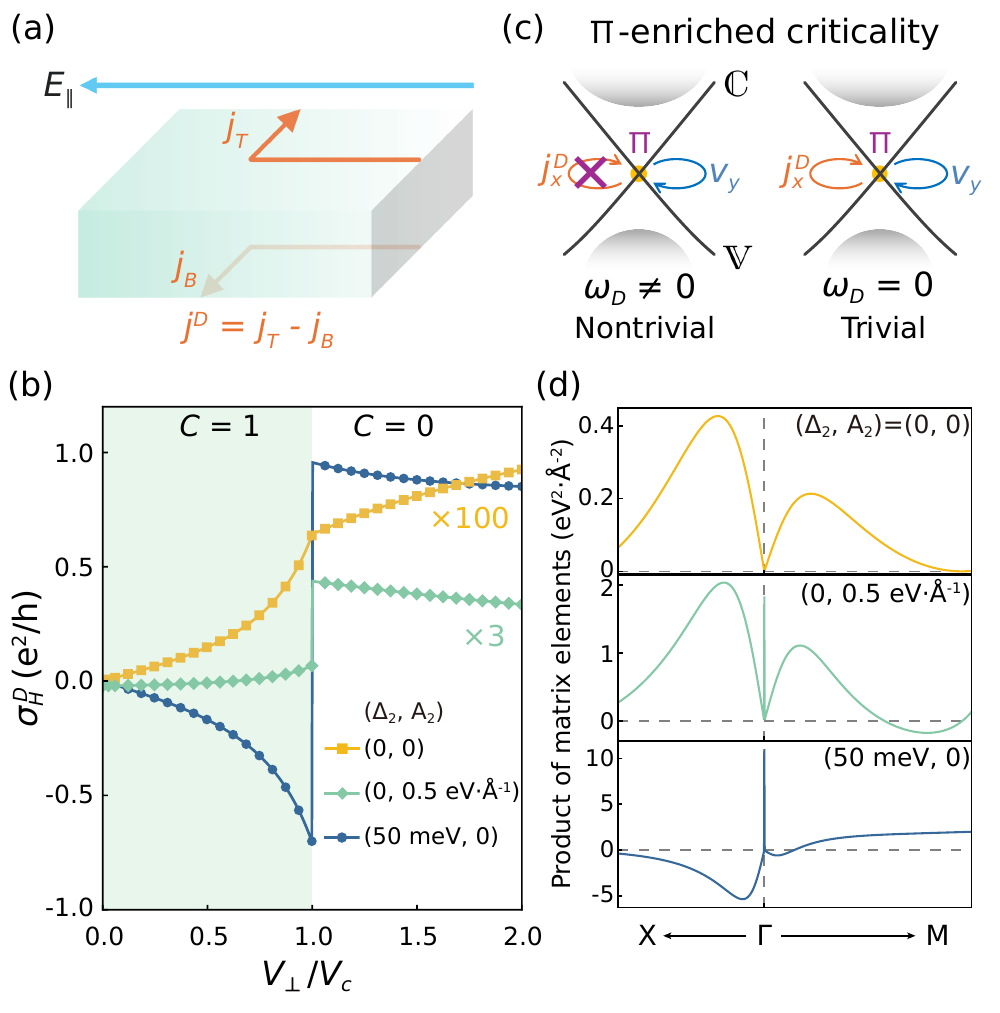}
\caption{\label{fig:1} Quasisymmetry-enriched continuous topological phase transition (CTPT) between Chern and normal insulating phases in BHZ model. (a) Schematic of pseudospin (dipole) Hall current $\bm{j}^D=\bm{j}_T-\bm{j}_B =\sigma_{H}^D(\hat{z}\times\bm{E}_{\parallel})$, describing charge Hall counter-flows $\bm{j}_T$ and $\bm{j}_B$ on top and bottom surfaces, induced by in-plane electric field $\bm{E}_\parallel$. (b) Dipole Hall conductivity $\sigma_{H}^{D}$ as a function of $V_\perp$, the energy bias between the surfaces.
The yellow, blue, and green symbols correspond to the exchange field $\Delta_2$ and velocity difference $A_2$ set as ($\Delta_2,A_2$) = (0, 0), (50 meV, 0), (0, 0.5 eV·\AA$^{-1}$) respectively. (c) Schematic of the gapless criticality enriched by quasisymmetry $\Pi$, which forbids the matrix element $\langle \mathds{V},\bm{k}|j_x^D|\mathds{C},\bm{k}\rangle$ at the touching point iff quasisymmetry charge $\omega_D \neq 0$. (d) Distributions of ${\rm Im}\left[\langle \mathds{V},\bm{k}|j_x^D|\mathds{C},\bm{k}\rangle \langle \mathds{C},\bm{k}|v_y|\mathds{V},\bm{k}\rangle \right]$ along high symmetry lines in Brillouin zone, for the BHZ model under $V_\perp = V_c$ and three set of ($\Delta_2,A_2$) as in (b). This quantity determines the continuity of $\sigma_{H}^{D}$ at the critical point [c.f. Eq.~(\ref{eq:limits_of_conductivity})].}
\end{figure}

\begin{figure}
\centering
\includegraphics[width=1\linewidth]{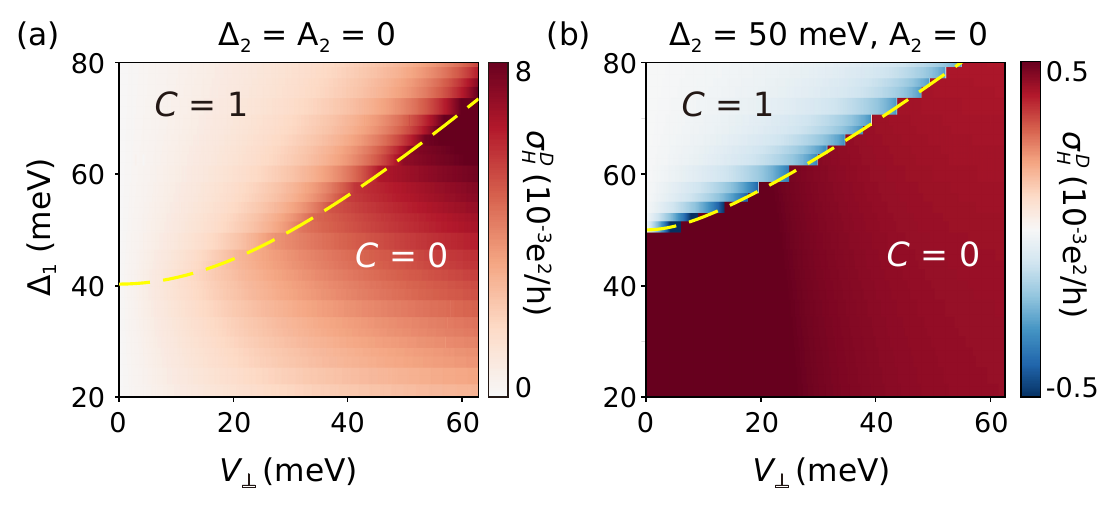}
\caption{\label{fig:2}Phase diagrams of BHZ-based CI model with $\Delta_{2}=A_{2}=0$ (a) and $\Delta_{2}=50$ meV, $A_2=0$ (b), and yellow dashed curves denote the phase boundaries. Color maps the magnitude of dipole Hall conductivity $\sigma_{H}^{D}$.}
\end{figure}

\textit{CTPT with intrinsic current correlations at criticality.}---We begin with CTPT described by the paradigmatic Bernevig-Hughes-Zhang (BHZ) model with broken time-reversal symmetry, which describes magnetic topological insulator thin films \cite{ZhangPRL2019TopologicalAxionStates}. The BHZ Hamiltonian reads~\cite{BernevigS2006QuantumSpinHall,LuPRB2010MassiveDiracfermions},
\begin{equation} 
H_{\text{BHZ}}(\bm{k})=\varepsilon_{0}\left(\bm{k}\right)+A(k_{x}s_{y}-k_{y}s_{x})\tau_{z}+M_{0}(\bm{k})\tau_{x},\label{eq:TI_H0} 
\end{equation} 
where $\varepsilon_{0}(\bm{k})=C_{0}-Dk^{2}$, $M_{0}(\bm{k})=M_0-Bk^{2}$. $s$ are the spin Pauli matrices, and $\tau$ denote the ones for the surface (layer) pseudospin, with $z$-component corresponding to an out-of-plane electric dipole $p_{z}=ed\tau_{z}$. $d$ is the film thickness. The rich phase diagram is accessed by adding symmetry-breaking terms $H^{\prime}=\frac{1}{2}(\Delta_{1}s_{z}+\Delta_{2}s_{z}\tau_{x}+V_{\perp}\tau_{z}) + A_2(k_x s_y-k_y s_x)$. $V_{\perp}$ is the energy bias between top and bottom surfaces. $\Delta_{1}$ and $\Delta_{2}$ represent magnetic exchange fields of two different forms, where $\Delta_{2}$ is the one to break quasisymmetry, as explained later. The $A_2$ term measures the velocity difference of two surface states due to the inversion symmetry breaking.
For $\left|\Delta_{1}\right|>\max(2\left|M\right|,\left|\Delta_{2}\right|)$, 
tuning $V_{\perp}$ drives a CTPT from a CI to a normal insulator (NI), with gap closure at the $\Gamma$ point. Fig.~\ref{fig:2} shows the phase diagrams at two $\Delta_2$ values, calculated by embedding the model in a square lattice with magnetic point group symmetry $4m^{\prime}m^{\prime}$. The parameters are listed in Appendix \hyperref[Appendix-A]{A}.

The unconventional nature of this CTPT manifests in the intrinsic dipole (pseudospin) Hall conductivity $\sigma_{H}^{D}$, which measures the response of the pseudospin current $\bm{j}^{D}=e\{\bm{v},\tau_{z}\}/2$ to an in-plane electric field $\bm{E}_{\parallel}$. $\bm{j}^{D}$ describes counter-flowing charge currents on the two surfaces. Within the linear response theory,  $\sigma_{H}^{D}$ is computed through the current correlation
\begin{equation} 
\sigma_{H}^{D}=\lim_{\omega\rightarrow0}\frac{1}{\hbar\omega}\int_{-\infty}^0 dt^{\prime}e^{i\omega t^{\prime}}\langle[j_x^D(0),j_y(t^{\prime})]\rangle,\label{eq:conductivity_correlation} 
\end{equation}
where $\omega$ is the frequency of $\bm{E}_{\parallel}$ and $\langle \cdots \rangle$ denotes thermodynamic average (c.f. Supplemental Material, Sec. S1 \cite{Supp}). At zero temperature limit, it can be put in a similar form as the charge Hall conductivity, 
\begin{equation} \label{eq:conductivity_curvature}
\sigma_{H}^{D}=\int\frac{d^{2}\bm{k}}{(2\pi)^{2}}\Omega^{D}(\bm{k}), 
\end{equation} 
where $\Omega^{D}(\bm{k})$ is the dipole Berry curvature~\cite{FanNC2024IntrinsicdipoleHall}, as a case of the generalized Berry curvature, 
\begin{equation} \label{eq:pseudospin_Berry_curvature}
\Omega^{\Lambda}(\bm{k})=2\hbar e\sum_{v,c}\frac{{\rm Im}\left[\langle v,\bm{k}|j_{x}^{\Lambda}|c,\bm{k}\rangle\langle c,\bm{k}|v_{y}|v,\bm{k}\rangle\right]}{(\varepsilon_{v,\bm{k}}-\varepsilon_{c,\bm{k}})^{2}}.
\end{equation} 
$|n,\bm{k}\rangle$ are the Bloch states with band energy $\varepsilon_{n,\bm{k}}$, and $v$ ($c$) indexes valence (conduction) bands. $j_{i}^{\Lambda}\equiv\{v_{i},\Lambda\}/2$ denotes the generalized current operator for various quantities (e.g., spin \cite{SinovaPRL2004UniversalIntrinsicSpin}, valley \cite{XiaoPRL2007ValleyContrastingPhysics}, charge multipoles \cite{FanNC2024IntrinsicdipoleHall,ZhengNL2024InterlayerElectricMultipoles})~\footnote{Our main results do not depend on the concrete construction of $j_{i}^{\Lambda}$, but depend on its momentum dependence projected down to the low-energy subspace.}. $\Omega^{\Lambda}(\bm{k})$ reduces to the usual charge Berry curvature $\Omega(\bm{k})$ and the dipole one $\Omega^{D}(\bm{k})$ by setting $\Lambda$ as the charge monopole and dipole operators, respectively. 

Both $\Omega(\bm{k})$ and $\Omega^{D}(\bm{k})$ become singular at the gap closure. However, while the charge Hall conductivity (Chern number $C$) has a quantized jump, our calculations reveal that $\sigma_{H}^{D}$ exhibits two types of behaviors at the gapless criticality, being: (i) discontinuous when $\Delta_{2} \neq 0$ or $A_{2}\neq 0$, like the charge Hall conductivity; or (ii) continuous when $\Delta_{2} = A_{2} = 0$ [Fig.~\ref{fig:1}(b)]. In the latter case, the \textit{gapless} criticality features \textit{an intrinsic correlation between charge current and dipole current} [Eq.~(\ref{eq:conductivity_correlation})], which is typically associated with \textit{gapped} phases. This behavior persists along the entire CI/NI phase boundary in the $\left(V_{\perp},\Delta_{1}\right)$ parameter space under $\Delta_{2} =A_{2} =0$ [Fig.~\ref{fig:2}(a)]. Such gapped phenomena of intrinsic current correlation in gapless criticality hints the quasisymmetry-protection of CTPT, underlying fundamentally distinguishable criticalities even when they belong to the same universality class of two-dimensional Dirac model with $p$-wave orbital symmetry~\cite{ChenPRB2017Correlationlengthuniversality,Supp}.

\textit{Microscopic requirement for the intrinsic current correlation at criticality.}---The continuity of $\sigma_{H}^{D}$ is traced to the integrability of the dipole Berry curvature. Upon CTPT at critical parameter $V_{c}$, the highest valence band $\mathds{V}$ and the lowest conduction band $\mathds{C}$ touch at momentum $\bm{k}_{0}$, resulting in a divergent contribution to the charge Berry curvature $\Omega(\bm{k}_{0})$, whose non-integrability is necessitated by the quantized jump of Chern number at criticality. For the generalized Berry curvature $\Omega^{\Lambda}$, however, the integrability at $\bm{k}_{0}$ need to be separately analyzed from the contribution by the pair of touching bands $(\mathds{V},\mathds{C})$:
\begin{equation}\label{eq:projected_pseudospin_Berry_1}
2\hbar e\frac{{\rm Im}\left[\langle\mathds{V},\bm{k}_{0}|j_{x}^{\Lambda}|\mathds{C},\bm{k}_{0}\rangle\langle\mathds{C},\bm{k}_{0}|v_{y}|\mathds{V},\bm{k}_{0}\rangle\right]}{(\varepsilon_{\mathds{V},\bm{k}_{0}}-\varepsilon_{\mathds{C},\bm{k}_{0}})^{2}}.
\end{equation}  
For the usual case of linear band touching, this is pinned down by the numerator: $\Omega^{\Lambda}(\bm{k}_{0})$ is integrable at criticality iff $\langle\mathds{V},\bm{k}_{0}|j_{x}^{\Lambda}|\mathds{C},\bm{k}_{0}\rangle\langle\mathds{C},\bm{k}_{0}|v_{y}|\mathds{V},\bm{k}_{0}\rangle = 0$.
We find the change of generalized Hall conductivity upon the CTPT (c.f. Appendix \hyperref[Appendix-B]{B}):
\begin{eqnarray}\label{eq:limits_of_conductivity}
|\delta\sigma_{H}^{\Lambda}|
&\equiv& |\lim_{V\rightarrow V_{c}^{+}}\sigma_{H}^{\Lambda}-\lim_{V\rightarrow V_{c}^{-}}\sigma_{H}^{\Lambda} | \\
&\propto&\lim_{V\rightarrow V_{c}}\big|{\rm Im}\left[\langle\mathds{V},\bm{k}_{0}|j_{x}^{\Lambda}|\mathds{C},\bm{k}_{0}\rangle\langle\mathds{C},\bm{k}_{0}|v_{y}|\mathds{V},\bm{k}_{0}\rangle\right]\big|. \notag
\end{eqnarray}
$\langle\mathds{C},\bm{k}_{0}|v_{x,y}|\mathds{V},\bm{k}_{0}\rangle$ must be finite to ensure a quantized change of Chern number. Meanwhile, for a general pseudospin $\Lambda$, $\langle\mathds{V},\bm{k}_{0}|j_{x,y}^{\Lambda}|\mathds{C},\bm{k}_{0}\rangle$ can vanish, resulting in a continuous pseudospin Hall conductivity $\sigma_{H}^{\Lambda}$ at criticality, as sketched in Fig.~\ref{fig:1}(c). 

\textit{Quasisymmetry enrichment.}---It is important to note that a vanishing $\langle\mathds{V},\bm{k}_{0}|j_{x,y}^{\Lambda}|\mathds{C},\bm{k}_{0}\rangle$ may not be due to the constraint by the little co-group $\mathcal{G}_{\bm{k}_{0}}=\{g|[g,H_{\bm{k}_{0}}]=0\}$ of the Hamiltonian at $\bm{k}_{0}$. 
If $\Lambda$ is invariant under $\mathcal{G}_{k_0}$, $\langle\mathds{V},\bm{k}_{0}|j_{x,y}^{\Lambda}|\mathds{C},\bm{k}_{0}\rangle$ transforms identically as the non-zero $\langle\mathds{V},\bm{k}_{0}|v_{x,y}|\mathds{C},\bm{k}_{0}\rangle$ does. 
Nonetheless, $\langle\mathds{V},\bm{k}_{0}|j_{x,y}^{\Lambda}|\mathds{C},\bm{k}_{0}\rangle$ can be prohibited by the higher symmetries of the subspace spanned by the pair of touching bands $(\mathds{V},\mathds{C})$
\cite{RenPRL2021QuasisymmetryGroupsMany}. Specifically, one could find extra symmetries $\Pi \notin \mathcal{G}_{k_0}$ that preserve 
the subspace $\{ |\mathds{V},\bm{k}_{0}\rangle,|\mathds{C},\bm{k}_{0}\rangle \}$, under which the matrix element must transform as,
\begin{equation}
\langle\mathds{V},\bm{k}_{0}|j_{x,y}^{\Lambda}|\mathds{C},\bm{k}_{0}\rangle\stackrel{\Pi}{\rightarrow}e^{i\omega_{\Lambda}}\langle\mathds{V},\bm{k}_{0}|j_{x,y}^{\Lambda}|\mathds{C},\bm{k}_{0}\rangle.\label{eq:quasisymmetry}
\end{equation}
Such $\Pi$ is termed a quasisymmetry of the subspace \cite{LiPRL2024GroupTheoryQuasisymmetry}, as it is not the exact symmetry of the Hamiltonian $H_{\bm{k}_{0}}$.
$\omega_{\Lambda} \in [0,2\pi)$ serves as the quasisymmetry charge of $\langle\mathds{V},\bm{k}_{0}|j_{x,y}^{\Lambda}|\mathds{C},\bm{k}_{0}\rangle$ under $\Pi$.
If the charge is nontrivial, i.e., $\omega_{\Lambda}\neq0$, the matrix element $\langle\mathds{V},\bm{k}_{0}|j_{x,y}^{\Lambda}|\mathds{C},\bm{k}_{0}\rangle$ is forbidden by this quasisymmetry. Oppositely, a trivial charge $\omega_{\Lambda}=0$ places no constraints on the matrix element.
For the special case of $\Lambda$ being the charge monopole, we always have a trivially zero $\omega$, so $\Pi$ can not impose additional constraint on the charge Hall conductivity. See Supplemental Material, Sec. S4 for details \cite{Supp}. 

In the presence of quasisymmetry, the classification of CTPT is enriched: CTPTs with trivial and nontrivial quasisymmetry charges $\omega_{\Lambda}$ are further distinguished, even if they share the same universality class and changes in Chern number. CTPTs with nonzero $\omega_{\Lambda}$ are distinctive due to the quasisymmetry-protected intrinsic current correlation and generalized Hall conductivity at gapless criticality, whereas those with zero $\omega_{\Lambda}$ behave as cases without quasisymmetry.

\textit{BHZ model revisit.}---Now we come back to the unconventional CTPT in BHZ model. The linear band touching occurs at $\Gamma$ where $|\mathds{V},\Gamma\rangle=(0,0,\cos\frac{\theta_{\downarrow}}{2},\sin\frac{\theta_{\downarrow}}{2})^{T}$ and $|\mathds{C},\Gamma\rangle=(-\sin\frac{\theta_{\uparrow}}{2},\cos\frac{\theta_{\uparrow}}{2},0,0)^{T}$ with $\tan\theta_{\uparrow/\downarrow}=(2M\pm\Delta_{2})/V_{\perp}$. When $\Delta_{2}=0$, $\theta_{\uparrow}$ and $\theta_{\downarrow}$ are identical, 
and one finds the operator $\Pi=s_{y}\sigma_{y}\mathcal{K}$, with $\mathcal{K}$ the complex conjugation,  interchanges the two states and preserves the subspace.
This is a quasisymmetry as $\Pi \notin \mathcal{G}_{\Gamma}$, and importantly it transforms (see Appendix \hyperref[Appendix-C]{C})
\begin{equation}
\langle\mathds{V},\Gamma|j_{x,y}^{D}|\mathds{C},\Gamma\rangle\stackrel{\Pi}{\rightarrow}-\langle\mathds{V},\Gamma|j_{x,y}^{D}|\mathds{C},\Gamma\rangle,\label{eq:quasisymmetry_BHZ}
\end{equation}
with a nontrivial charge $\omega_{D}=\pi$, which requires the matrix element to vanish. 
Eq.~(\ref{eq:limits_of_conductivity}) then dictates the continuous $\sigma_{H}^{D}$ at the gapless criticality [Fig.~\ref{fig:1}(c)]. 
Such unconventional CTPT behavior is protected against any perturbation commuting with $\Pi$.
Indeed $\sigma_{H}^{D}$ is found to be continuous over the entire phase boundary in the $(V_{\perp},\Delta_{1})$ parameter space under $\Delta_2=A_{2}=0$ [Fig.~\ref{fig:2}(a)].

With finite $A_2$, although $\Pi$ is still a quasisymmetry, its charge becomes trivial $\omega_D=0$, thus it plays no rules on $\langle\mathds{V},\Gamma|j_{x,y}^{D}|\mathds{C},\Gamma\rangle$. Under a finite exchange field $\Delta_{2}s_{z}\sigma_{x}$, one can no longer find any quasisymmetry to restrict $\langle\mathds{V},\Gamma|j_{x,y}^{D}|\mathds{C},\Gamma\rangle$. In both cases, $\sigma_{H}^{D}$ indeed becomes discontinuous [c.f. Eq.~(\ref{eq:A_DHE})]
\begin{equation}
\left|\delta\sigma_{H}^{D}\right|=\frac{e^{2}}{8\pi\hbar}\frac{V_{c}\Delta_{2}}{\left|M_0\right|\sqrt{4M_0^{2}+V_{c}^{2}}} +  \frac{e^2}{2\pi\hbar}\frac{A_2}{A},\label{eq:TI_DHE}
\end{equation}
invalidating the intrinsic current correlation at gapless criticality [Fig.~\ref{fig:2}(b)]. Overall, the CTPT in BHZ model is enriched by the quasisymmetry $\Pi$, explaining the intrinsic correlation between charge and dipole currents. 

\begin{figure}
\centering
\includegraphics[width=1\columnwidth]{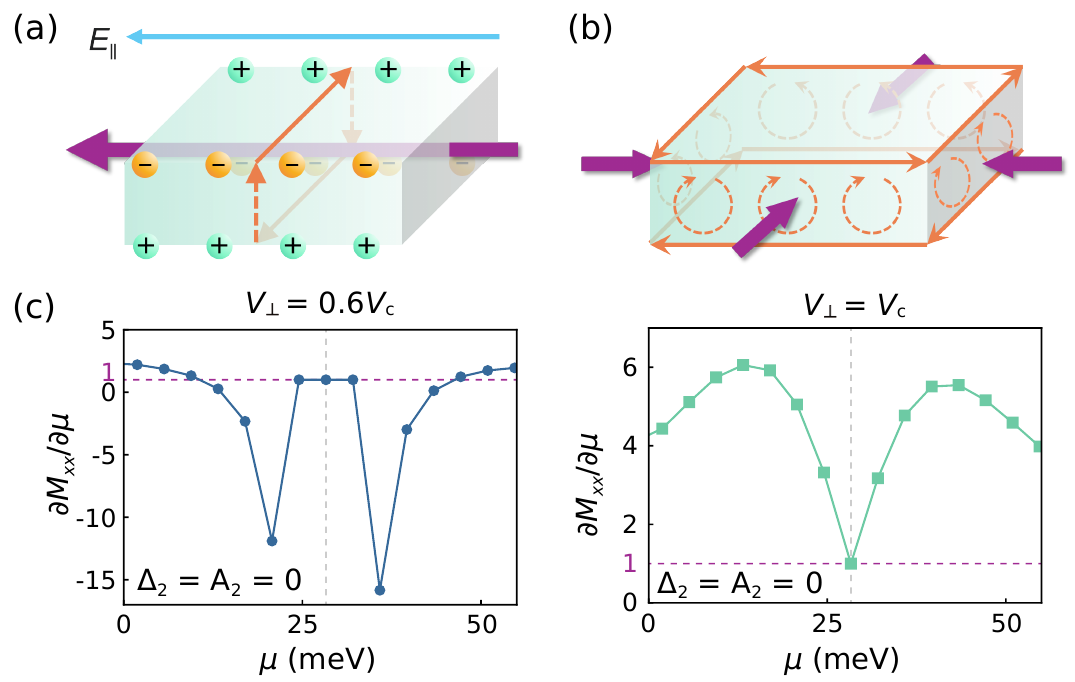}
\caption{\label{fig:3} Generalized  Středa formula linking dipole Hall conductivity and equilibrium magnetic quadrupole order in gapped phase and gapless criticality. (a) Dipole Hall effect as a magnetoelectric response. The counter-flowing Hall currents on the two surfaces and boundary tunneling currents form a current loop that correspond to an orbital magnetization longitudinal to the in-plane electric field. (b) In equilibrium, the counter-flowing edge currents on the top and bottom surfaces—an edge dipole current—can be equivalently viewed as an array of current loops (dashed-arrow circles). The resulting orbital magnetization, oriented normal to the side surfaces, realizes a magnetic quadrupole order. 
(c) $\partial M_{xx}/\partial \mu$ as a function of chemical potential $\mu$, counting both Fermi sea and Fermi surface contributions (c.f. Appendix \hyperref[Appendix-D]{D}). 
$V_\perp$ is the energy bias between the top and bottom surfaces as the tuning parameter for the CTPT, $V_c$ being the critical value. Left: $V_\perp = 0.6 V_c$, the system is a Chern insulator and the generalized Středa formula manifests as a constant $\partial M_{xx}/\partial \mu$ in the gap. Right: $V_\perp = V_c$, the gapped formula is still applicable at the gapless criticality by the quasisymmetry protection. Dashed vertical line indicates middle of the gap, and vertical axis is in unit of $-\sigma_H^Dd/2e$, the dipole Hall conductivity in gap.  }
\end{figure}

\textit{Středa formula at criticality.}---The quasisymmetry protection of the CTPT further leads to a generalized Středa formula, as another manifestation of gapped phenomena in gapless criticality. 
In insulators the standard Středa formula relates the quantized charge Hall conductivity to orbital magnetization $\bm{M}$: $\sigma_{H}=e\partial M_{z}/\partial\mu$, $\mu$ being the chemical potential in the gap~\cite{StredaJoPCSSP1982TheoryquantisedHall,XiaoPRL2005BerryPhaseCorrection,CeresoliPRB2006Orbitalmagnetizationcrystalline}. It breaks down at gapless criticality of CTPT where $\sigma_{H}$ and Fermi-surface contribution to $\bm{M}$ are ill-defined.
For the quasisymmetry-protected CTPT, however, a continuous pseudospin Hall conductivity $\sigma_{H}^{\Lambda}$ makes relevant a generalized Středa formula at the gapless criticality.

We start from the insulating regimes, where the dipole Hall conductivity was shown to be equivalent to the orbital magnetoelectric response $\partial M_{x}/\partial E_{\parallel,x}=\sigma_{H}^{D}d/2$~\cite{FanNC2024IntrinsicdipoleHall}. This has an intuitive picture: the in-plane field $E_{\parallel,x}$ induces counter-flowing charge Hall currents on the two surfaces [Fig.~\ref{fig:3}(a)], which leads to dipole accumulations on the boundaries that relax through inter-surface tunneling; 
the boundary tunneling currents and the surface counter-flows form a current loop producing an in-plane orbital magnetization $M_x$~\cite{FanNC2024IntrinsicdipoleHall}.
In insulators, this response can be further related to the equilibrium orbital magnetic quadrupole moment (MQM) $M_{ij} = \int r_{i}\mathcal{M}_{j}(\bm{r})d\bm{r} / (3V)$ through $-e \partial M_{xx}/\partial \mu = \partial M_{x}/\partial E_{\parallel,x}$~\cite{GaoPRB2018Orbitalmagneticquadrupole,ShitadePRB2018Theoryorbitalmagnetic}, where $\bm{\mathcal{M}}(\bm{r})$ is the local magnetization density and $V$ is the system volume. These together lead to the generalized Středa formula linking MQM to dipole Hall conductivity,
\begin{equation}
-e \frac{\partial M_{xx}}{\partial \mu} =-e \frac{\partial M_{yy}}{\partial \mu} = \frac{d}{2}\sigma_{H}^{D}. \label{eq:Streda_formula_1}
\end{equation}
And we note that $M_{xy}$ is symmetry-forbidden here. 

The equilibrium MQM comes from counter-flowing edge currents at the top and bottom surfaces, or equivalently an edge current of electric dipole [Fig.~\ref{fig:3}(b)]. This gives rise to a distribution of in-plane orbital magnetization along the edge: $\bm{m}=\frac{1}{4}(\bm{p}\times \bm{v}-\bm{v}\times \bm{p})$ \cite{FanNC2024IntrinsicdipoleHall}, which can constitute the MQM both in CI and NI phases. In the latter case, the $\mu$-dependence of $M_{ij}$ originates from the itinerant circulation of edge Wannier states \cite{ThonhauserPRL2005OrbitalMagnetizationPeriodic,CeresoliPRB2006Orbitalmagnetizationcrystalline,ShitadePRB2018Theoryorbitalmagnetic}, whereas in the former case the chiral edge states also contribute.

At the gapless criticality, while the intrinsic $\sigma_{H}^{D}$ is continuous under quasisymmetry protection, in general one needs to account extra Fermi surface contribution to $ -e \partial M_{xx}/\partial \mu$ (Appendix \hyperref[Appendix-D]{D}):
\begin{align}
\int \frac{d^2 \bm{k}}{(2\pi)^2} \frac{\hbar ed}{3}\sum_{n\neq m}\frac{\text{Im}\left[\left\langle n,\bm{k}\right|j_{x}^{D}\left|m,\bm{k}\right\rangle \left\langle m,\bm{k}\right|v_{y}\left|n,\bm{k}\right\rangle \right]}{\varepsilon_{n,\bm{k}}-\varepsilon_{m,\bm{k}}} \frac{\partial f_n }{\partial \varepsilon_{n,\bm{k}}}. \label{eq:Streda_formula_2}
\end{align}
At CTPT with $\mu$ crossing the band touching, the above integral is given by the contribution of $(\mathds{V},\mathds{C})$ at the touching $\bm{k}_0$. With the constraint imposed by quasisymmetry $\Pi$ on the numerator [Eq.~(\ref{eq:quasisymmetry_BHZ})], such a Fermi surface contribution vanishes exactly. Consequently, the otherwise gapped Středa formula Eq.~(\ref{eq:Streda_formula_1}) is inherited by the gapless criticality in quasisymmetry-enriched CTPT, as confirmed by numerical calculations in the BHZ model [see Fig.~\ref{fig:3}(c)].

\begin{figure}
\centering
\includegraphics[width=1\columnwidth]{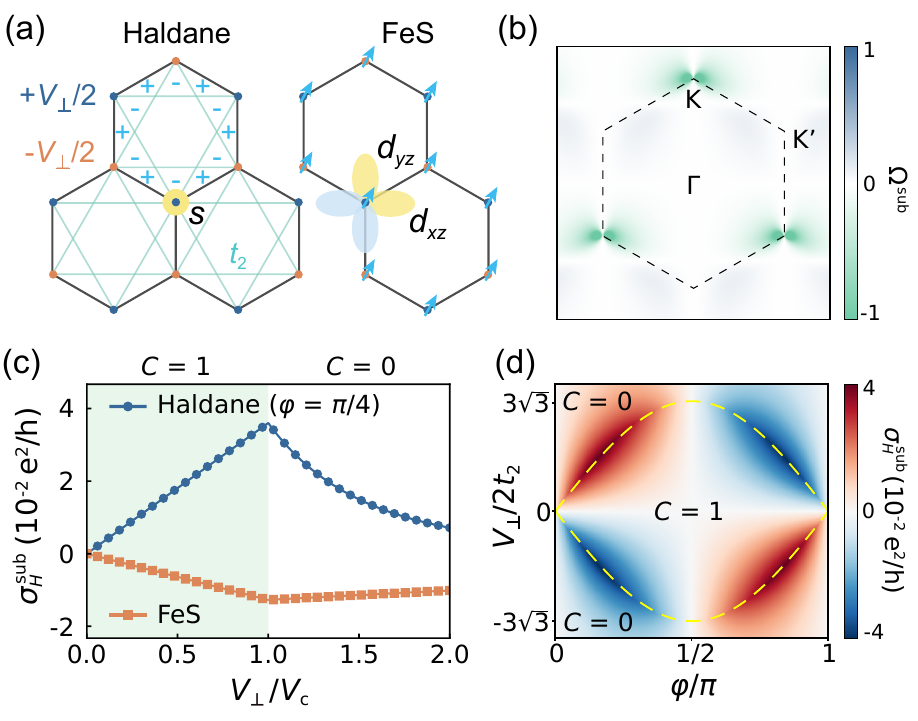}
\caption{\label{fig:4}Quasisymmetry-enriched CTPT in other Chern insulator systems. (a) Schematics of the Haldane model composed of $s$ orbital located at honeycomb lattice with staggered magnetic flux (marked by + and -), and FeS monolayer with $d_{xz}$ and $d_{yz}$ orbitals with ferromagnetic order (marked by arrows). Alternating on-site energy $\pm V_{\perp}/2$ are induced in both models, the $t_{2}$ denotes next nearest neighbor hopping integral. (b) Distribution of sublattice Berry curvature $\Omega_{xy}^{\rm{sub}}$ in Haldane model. (c) Sublattice Hall conductivity $\sigma_{xy}^{\rm{sub}}$ as a function of $V_{\perp}$ for Haldane model with $\varphi=\pi/4$ and FeS monolayer model. (d) Phase diagram of the Haldane model under variations of $V_{\perp}$ and $\varphi$, color coded by the value of $\sigma_{H}^{\rm{sub}}$. Dashed curves denote phase boundary.}
\end{figure}

\textit{Haldane model.}---Our next example is the Haldane model \cite{HaldanePRL1988ModelQuantumHall} consisting $s$ orbital located at a honeycomb lattice [Fig.~\ref{fig:4}(a), left panel]. With the staggered magnetic flux, hopping from sublattice A (B) to next nearest neighbor sites acquires a phase factor $e^{i\varphi}$ ($e^{-i\varphi}$). Alternating on-site energy $\pm V_{\perp}/2$ biases the sublattices and trivializes the CI phase by closing the gap at $K$ or $K^{\prime}$. Here we consider $\Lambda$ being the sublattice pseudospin, and examine a sublattice Hall current $\bm{j}_{A}-\bm{j}_{B}=\sigma_{H}^{\rm{sub}}\hat{z}\times\bm{E}_{\parallel}$. As shown in Fig.~\ref{fig:4}(a), $\sigma_{H}^{\rm{sub}}$ also remains continuous upon CTPT, revealing its unconventional nature with intrinsic current correlation at the gapless criticality.

The gap closing states are fully sublattice-polarized: $|\mathds{V},K\rangle=(0,1)^{T}$, $|\mathds{C},K\rangle=(1,0)^{T}$. This subspace is preserved by $\Pi^{\prime}=i\sigma_{x}\mathcal{K}$, which transforms $\left\langle \mathds{V},K\right|j_{x,y}^{{\rm sub}}\left|\mathds{C},K\right\rangle \stackrel{\Pi^{\prime}}{\rightarrow}-\left\langle \mathds{V},K\right|j_{x,y}^{{\rm sub}}\left|\mathds{C},K\right\rangle$. The intrinsic current correlation is protected by quasisymmetry $\Pi^{\prime}$ along the entire phase boundary of CTPT [Fig.~\ref{fig:4}(d)], even when longer range hopping terms are included (Supplemental Material, Sec. S7 \cite{Supp}).

We also analyze a four-band version ($d_{xz}$ and $d_{yz}$ orbitals) of flux-free Haldane model, which describes CI phase in hexagonal lattice of transition metal elements, e.g., monolayer FeS \cite{LiNC2022Designinglightelement}. As shown in Fig.~\ref{fig:4}(c), $\sigma_{H}^{{\rm sub}}$ is also continuous at criticality. This constitutes another example of quasisymmetry-enriched CTPT.

\textit{Discussion}.---
The continuous $\sigma_{H}^{\Lambda}$ at criticality by quasisymmetry protection essentially relies on Eq.~(\ref{eq:limits_of_conductivity}), which has assumed linear band crossing. This is a most common situation for discrete band touching in crystalline systems, unless the little co-group $\mathcal{G}_{\bm{k}_{0}}$ at touching point includes four- or six-fold rotation symmetries, where bands may touch with parabolic or cubic dispersions \cite{FangPRL2012MultiWeylTopological}. For these situations, Eq.~(\ref{eq:limits_of_conductivity}) fails and the singularity is contributed by a neighborhood of $\bm{k}_{0}$ rather than a single touching point, such that the quasisymmetry emerging at $\bm{k}_{0}$ cannot guarantee the integrability of generalized Berry curvature. For other cases where gap-closing momenta form a nodal line or plane, quasisymmetry may still protect the continuity of $\sigma_{H}^{\Lambda}$ once the curial quasisymmetry emerges at the entire nodal line or surface \cite{GuoNP2022Quasisymmetryprotected,HuPRB2023Hierarchyquasisymmetriesdegeneracies}. 
Lastly, we note that this framework can be directly generalized to CTPTs characterized by other topological invariants arising from geometric quantities, for example, the mirror Chern number describing mirror-symmetric topological crystalline insulators~\cite{AndoARoCMP2015TopologicalCrystallineInsulators}. 

\textit{Acknowledgement}.---J. Li thanks Haiyuan Zhu, Haowei Chen, Zhijian Song, Qihang Liu, and Qian Niu for helpful discussions. The work is supported by National Natural Science Foundation of China (No. 12425406), Research Grant Council of Hong Kong SAR (HKU SRFS2122-7S05, AoE/P-701/20, A-HKU705/21), and New Cornerstone Science Foundation.

\bibliography{ref}

\clearpage{}

\section*{Appendix}

\renewcommand{\theequation}{A\arabic{equation}}
\setcounter{equation}{0}

\textit{Appendix A.}\label{Appendix-A}---Parameters for the BHZ model are chosen to represent the four-band Hamiltonian of three quintuple layers Bi$_{2}$Se$_{3}$ \cite{LuPRB2010MassiveDiracfermions}: $M_0=-0.0201431\ {\rm eV}$, $C_{0}=0.02866\ {\rm eV}$, $A=4.06448\ {\rm eV}\cdot\text{\AA}^{-1}$, $B=-19.9678\ {\rm eV}\cdot\text{\AA}^{-2}$, $D=-12.659\ {\rm eV}\cdot\text{\AA}^{-2}$, and $d=29.829\ \text{\AA}$. Exchange field $\Delta_{1}=60\ {\rm meV}$ is fixed to achieve CI phase with $C=1$ at half filling. $V_{\perp}$ trivializes the CI by closing the gap at $\Gamma$ with critical field
\begin{equation}
V_{c}=\sqrt{\Delta_{1}^{2}-4M_0^{2}}\sqrt{1-\left(\Delta_{2}/\Delta_{1}\right)^{2}}.\label{eq:A_TI_TPT}
\end{equation}

Around $\Gamma$, the effective two-band Hamiltonian reads $H_{\bm{k}}^{\mathds{VC}}=C^{\prime}\left(\bm{k}\right) -D^{\prime}k^2+A^{\prime}\left(k_{x}\tau_{y}+k_{y}\tau_{x}\right)+\left(M_0^{\prime}-B^{\prime}k^{2}\right)\tau_{z}$. The primed parameters are: $C^{\prime} = C+\frac{1}{2}\left(F_{\downarrow}-F_{\uparrow}\right)$, $D^{\prime} = D+\frac{B}{2}\left(\sin\theta_{\downarrow}-\sin\theta_{\uparrow}\right)$, $A^{\prime} = A\sin\frac{\theta_{\uparrow}+\theta_{\downarrow}}{2}$, $M_0^{\prime}=-\frac{\Delta_{1}}{2}+\frac{1}{2}\left(F_{\downarrow}+F_{\uparrow}\right)$, and $B^{\prime} = \frac{B}{2}\left(\sin\theta_{\downarrow}+\sin\theta_{\uparrow}\right)$, with $\tan\theta_{\uparrow,\downarrow}=\left(2M_0 \pm\Delta_{2}\right)/V_{\perp}$ and $F_{\uparrow,\downarrow}=\frac{1}{2}\sqrt{\left(2M_0 \pm \Delta_{2}\right)^{2}+V_{\perp}^{2}}$. 

The low energy dispersions along the TPT boundary are the linear function of momenta. The universality class of criticality is characterized by critical phenomena, such as the critical exponent $\nu$ of the divergent correlation length $\xi\propto|V_\perp-V_c|^{-\nu}$, and dynamical exponent $z$. In both cases with and without $\Delta_2$ and $A_2$, the criticalities belong to the universality class of the two-dimensional Dirac model with $p$-wave orbital symmetry, with $z=1$ and $\nu=1$ \cite{ChenPRB2017Correlationlengthuniversality}. See Supplemental Material, Sec. S5 for details \cite{Supp}.

Up to the linear order of $\Delta_{2}$ and $A_2$, the dipole Berry curvature of $H_{\bm{k}}^{\mathds{VC}}$ reads $\Omega^{D}\left(\bm{k}\right)=\Omega_{0}^{D}\left(\bm{k}\right)+\Omega_{\Delta}^{D}\left(\bm{k}\right)+\Omega_{A}^{D}\left(\bm{k}\right)$, where
\begin{align}
\Omega_{0}^{D}\left(\bm{k}\right) & =\frac{e^{2}}{2\hbar}\frac{\mathcal{D}\mathcal{A}^{2}k_x^2}{\left[\mathcal{A}^{2}k^{2}+\left(\mathcal{M}_0-\Delta_{1}/2-\mathcal{B}^{2}k^{2}\right)^{2}\right]^{3/2}},\label{eq:A_curvature_0}\\
\Omega_{\Delta}^{D}\left(\bm{k}\right) & =\frac{e^{2}}{16\hbar}\frac{\Delta_{2}\mathcal{A}^{2}V_{\perp}\zeta}{\mathcal{M}_0^{2}}\nonumber \\
 & \times\frac{\mathcal{M}_0-\Delta_{1}/2-\mathcal{B}\left(k_{x}^{2}-k_{y}^{2}\right)}{\left[\mathcal{A}^{2}k^{2}+\left(\mathcal{M}_0-\Delta_{1}/2-\mathcal{B}k^{2}\right)^{2}\right]^{3/2}},\label{eq:A_curvature_1} \\
\Omega_{A}^{D}\left(\bm{k}\right) &=-\frac{e^{2}}{2\hbar}\frac{A_{2}\mathcal{A}\left[\mathcal{M}_0-\Delta_{1}/2-\mathcal{B}\left(k_{x}^{2}-k_{y}^{2}\right)\right]}{\zeta\left[\mathcal{A}^{2}k^{2}+\left(\mathcal{M}_0-\Delta_{1}/2-\mathcal{B}k^{2}\right)^{2}\right]^{3/2}}.\label{eq:A_curvature_2}
\end{align}
Here, $\zeta=\sqrt{1+V_{\perp}^{2}/\left(2M_0\right)^{2}}$, $\mathcal{A}=A/\zeta$, $\mathcal{B}=B/\zeta$, $\mathcal{M}_0=M_0\zeta$, and $\mathcal{D}=DV_\perp/2\mathcal{M}_0$. The conductivity $\sigma_{H}^{D}$ reads
\begin{equation}
\frac{e^{2}}{8\pi \hbar}\frac{V_{\perp}}{2\left|M_0\right|}\left[\frac{DS_{{\rm c}}\left(V_{\perp}\right)}{B}-\frac{\Delta_{2}S_{{\rm dc}}\left(V_{\perp}\right)}{\sqrt{4M_0^{2}+V_{\perp}^{2}}}\right]+\frac{e^2}{4\pi \hbar}\frac{A_2}{A}S_{\rm{dc}}(V_\perp),\label{eq:A_DHE}
\end{equation}
where $S_{\rm{c}}(V_\perp)$ is a continuous function at $V_{c}$, $\lim_{V_\perp\rightarrow V_{c}^{\pm}}S_{\rm{c}}=1$, and $S_{\rm{dc}}(V_\perp)$ is discontinuous, $\lim_{V_\perp\rightarrow V_{c}^{\pm}}S_{\rm{dc}}=\mp1$. Eq.~(\ref{eq:TI_DHE}) is obtained by taking the left and right limits of Eq.~(\ref{eq:A_DHE}). 

\renewcommand{\theequation}{B\arabic{equation}}
\setcounter{equation}{0}

\textit{Appendix B}.\label{Appendix-B}---The integrability of $\Omega^{\Lambda}\left(\bm{k}_{0}\right)$ is determined by the generalized Berry curvature of the effective two-band Hamiltonian $H_{\bm{k}}^{\mathds{VC}}=\mathcal{P}H_{\bm{k}}\mathcal{P}$ with $\mathcal{P}=\left|\mathds{V},\bm{k}_{0}\right\rangle \left\langle \mathds{V},\bm{k}_{0}\right|+\left|\mathds{C},\bm{k}_{0}\right\rangle \left\langle \mathds{C},\bm{k}_{0}\right|$. $H_{\bm{k}}^{\mathds{VC}}$ can be expressed in terms of Pauli matrices $\bm{\tau}$ as 
\begin{equation}
H_{\bm{k}}^{\mathds{VC}}=d_{0}\left(\bm{k}\right)+\bm{d}\left(\bm{k}\right)\cdot\bm{\tau},
\end{equation}
where $d_{0}\left(\bm{k}\right)=C-Dk^{2}$. Up to the irrelevant anisotropy and unitary transformation, $\bm{d}\left(\bm{k}\right)=\left(Ak_{x},Ak_{y},\Delta_{0}-Bk^{2}\right)$. Here, $A$ and $B$ are continuous at criticality, and $\Delta_{0}$ is the band gap at $\bm{k}_{0}$ such that $\lim_{V\rightarrow V_{c}^{\mp}}\Delta_{0}=0^{\pm}$. The dispersion is $\varepsilon_{\pm}=d_{0}(\bm{k})\pm\sqrt{A^{2}k^{2}+\left(\Delta_{0}-Bk^{2}\right)^{2}}$. For $H_{\bm{k}}^{\mathds{VC}}$, $\Omega^{\Lambda}\left(\bm{k}\right)$ can be expressed as (Supplemental Material, Sec. S2 \cite{Supp})
\begin{equation}
\Omega^{\Lambda}\left(\bm{k}\right)=-\hbar e\frac{\left[\partial_{x}\bm{d}^{\Lambda}(\bm{k})\times\partial_{y}\bm{d}(\bm{k})\right]\cdot\bm{d}(\bm{k})}{2|\bm{d}(\bm{k})|^{3}},\label{eq:B_Omega_VC}
\end{equation}
where $\bm{d}^{\Lambda}\left(\bm{k}\right)=\left(d_{x}^{\Lambda},d_{y}^{\Lambda},d_{z}^{\Lambda}\right)\left(\bm{k}\right)$ are defined through $\frac{1}{2}\mathcal{P}\left\{ \Lambda,H_{\bm{k}}\right\} \mathcal{P}=d_{0}^{\Lambda}\left(\bm{k}\right)+\bm{d}^{\Lambda}\left(\bm{k}\right)\cdot\bm{\tau}$. $\left[\partial_{x}\bm{d}^{\Lambda}\left(\bm{k}\right)\times\partial_{y}\bm{d}\left(\bm{k}\right)\right]\cdot\bm{d}\left(\bm{k}\right)/2$ can be parameterized as (Supplemental Material, Sec. S3 \cite{Supp})
\begin{equation}
\Delta_{0}\sum_{\substack{w=0,1\\
\ell=0,\dots,w
}
}R_{w,\ell}k_{x}^{\ell}k_{y}^{w-\ell}+\sum_{\substack{w=1,2,3\\
\ell=0,\dots,w
}
}T_{w,\ell}k_{x}^{\ell}k_{y}^{w-\ell},\label{eq:B_parameterization}
\end{equation}
where $R_{w,\ell}$ and $T_{w,\ell}$ are real parameters that are continuous across CTPT, and the factor $\hbar e$ can be absorbed in the parameters. We found that 
\begin{equation}
\lim_{V\rightarrow V_{c}^{\mp}}\sigma_{H}^{\Lambda}=\lim_{V\rightarrow V_{c}} \left(\frac{T_{2,0} + T_{2,2}}{4\pi BA^{2}}\pm\frac{ R_{0,0}}{2\pi A^{2}} \right),\label{eq:B_projected_conductivity}
\end{equation}
and thus, 
\begin{equation}
\left|\delta\sigma_{H}^{\Lambda}\right|=\lim_{V\rightarrow V_{c}}\left|\frac{R_{0,0}}{\pi A^{2}}\right|.
\end{equation}
$A$ is nonzero for linear band crossing. Comparing Eqs. (\ref{eq:projected_pseudospin_Berry_1}), (\ref{eq:B_Omega_VC}), and (\ref{eq:B_parameterization}), we find that 
\begin{equation}
\left|{\rm Im}\left[\left\langle \mathds{V},\bm{k}_{0}\right|j_{x}^{\Lambda}\left|\mathds{C},\bm{k}_{0}\right\rangle \left\langle \mathds{C},\bm{k}_{0}\right|v_{y}\left|\mathds{V},\bm{k}_{0}\right\rangle \right]\right|=\frac{\left|R_{0,0}\right|}{2}, \label{eq:B_R00}
\end{equation}
leading to Eq.~(\ref{eq:limits_of_conductivity}). 

\renewcommand{\theequation}{C\arabic{equation}}
\setcounter{equation}{0}

\begin{table}
\caption{\label{tab:C_character_table}Character table for corepresentations
of $4m^{\prime}m^{\prime}$. Only unitary symmetry operators are listed.
$\bar{E}$ denotes $2\pi$ rotation along any spatial axis, and $\omega=e^{i\pi/4}$.}
\centering{}%
\begin{tabular}{ccccccccc}
\hline 
 & $E$ & $C_{2z}$ & $C_{4z}^{+}$ & $C_{4z}^{-}$ & $\bar{E}$ & $\bar{E}C_{2z}$ & $\bar{E}C_{4z}^{+}$ & $\bar{E}C_{4z}^{-}$\tabularnewline
\hline 
$\Gamma_{1}$ & $1$ & $1$ & $1$ & $1$ & $1$ & $1$ & $1$ & $1$\tabularnewline
$\Gamma_{2}$ & $1$ & $1$ & $-1$ & $-1$ & $1$ & $1$ & $-1$ & $-1$\tabularnewline
$\Gamma_{3}$ & $1$ & $-1$ & $i$ & $-i$ & $1$ & $-1$ & $i$ & $-i$\tabularnewline
$\Gamma_{4}$ & $1$ & $-1$ & $-i$ & $i$ & $1$ & $-1$ & $-i$ & $i$\tabularnewline
$\bar{\Gamma}_{5}$ & $1$ & $-i$ & $i\omega$ & $-\omega$ & $-1$ & $i$ & $-i\omega$ & $\omega$\tabularnewline
$\bar{\Gamma}_{6}$ & $1$ & $-i$ & $-i\omega$ & $\omega$ & $-1$ & $i$ & $i\omega$ & $-\omega$\tabularnewline
$\bar{\Gamma}_{7}$ & $1$ & $i$ & $-\omega$ & $i\omega$ & $-1$ & $-i$ & $\omega$ & $-i\omega$\tabularnewline
$\bar{\Gamma}_{8}$ & $1$ & $i$ & $\omega$ & $-i\omega$ & $-1$ & $-i$ & $-\omega$ & $i\omega$\tabularnewline
\hline 
\end{tabular}
\end{table}

\textit{Appendix C}.\label{Appendix-C}---In the BHZ model with symmetry-breaking term $H^{\prime}$, the little co-group at $\Gamma$ is $\mathcal{G}_{\Gamma}=4m^{\prime}m^{\prime}$. $v_{x}$ and $v_{y}$ furnish a reducible corepresentation (denoted by $\sim$ hereafter) $\Gamma_{3}\oplus\Gamma_{4}$ (Tab. \ref{tab:C_character_table}), charge dipole $p_{z}=ed\tau_{z}\sim\Gamma_{1}$, $\left|\mathds{V},\Gamma \right\rangle \sim\bar{\Gamma}_{6}$, and $\left|\mathds{C},\Gamma\right\rangle \sim\bar{\Gamma}_{8}$. Thus, the corepresentation of $\left\langle \mathds{V},\Gamma\right|j_{x}^{\Lambda}\left|\mathds{C},\Gamma\right\rangle $ is
\begin{equation}
\bar{\Gamma}_{6}^{*}\otimes\Gamma_{1}\otimes\left(\Gamma_{3}\oplus\Gamma_{4}\right)\otimes\bar{\Gamma}_{8}=\Gamma_{2}\oplus\Gamma_{1}\ni\Gamma_{1},
\end{equation}
with $\Gamma_1$ the identity corepresentation. This means that $\left\langle \mathds{V},\bm{k}_{0}\right|j_{x}^{\Lambda}\left|\mathds{C},\bm{k}_{0}\right\rangle $ is $\mathcal{G}_{\Gamma}$-allowed. 

With the operator $\Pi = s_y\tau_y\mathcal{K}$ defined, the dipole current operator $j_x^D(\Gamma) = eAs_y/\hbar$ and velocity operator $v_y(\Gamma) = -As_{x}\tau_{z}/\hbar$ are transformed as $\Pi j_{x}^D \Pi^{-1} = - j_{x}^D$ and $\Pi v_{y} \Pi^{-1} = v_{y}$. As $\Pi$ maps $|\mathds{V},\Gamma \rangle \leftrightarrow |\mathds{C},\Gamma \rangle$, the transformation of the matrix elements are $\left\langle \mathds{V},\Gamma\right|j_{x}^{D}\left|\mathds{C},\Gamma\right\rangle \stackrel{\Pi}{\rightarrow} - \left\langle \mathds{V},\Gamma\right|j_{x}^{D}\left|\mathds{C},\Gamma\right\rangle $ and $\left\langle \mathds{C},\Gamma\right|v_{y}\left|\mathds{V},\Gamma\right\rangle  \stackrel{\Pi}{\rightarrow} \left\langle \mathds{C},\Gamma\right|v_{y}\left|\mathds{V},\Gamma\right\rangle$, resulting in Eq.~(\ref{eq:quasisymmetry_BHZ}). 

Under finite $A_2$, the dipole current operator and velocity operator become $j_x^D(\Gamma) = eAs_y/\hbar + eA_2s_y\tau_z/\hbar$ and $v_y(\Gamma) = -As_{x}\tau_{z}/\hbar -A_{2}s_{x}/\hbar$. Although $\Pi$ remains a quasisymmetry, it transforms $\left\langle \mathds{V},\Gamma\right|j_{x}^{D}\left|\mathds{C},\Gamma\right\rangle \stackrel{\Pi}{\rightarrow}  \left\langle \mathds{V},\Gamma\right|j_{x}^{D}\left|\mathds{C},\Gamma\right\rangle $ and $\left\langle \mathds{C},\Gamma\right|v_{y}\left|\mathds{V},\Gamma\right\rangle  \stackrel{\Pi}{\rightarrow} \left\langle \mathds{C},\Gamma\right|v_{y}\left|\mathds{V},\Gamma\right\rangle$. This means the quasisymmetry charge is trivially zero, and the matrix element is not constrained by $\Pi$. Detailed symmetry analysis is provided in Supplemental Material, Sec. S5 \cite{Supp}.

\renewcommand{\theequation}{D\arabic{equation}}
\setcounter{equation}{0}

\textit{Appendix D}.\label{Appendix-D}---The derivative of the orbital MQM $M_{kj}$ with respect to the chemical potential $\mu$ is given by \cite{GaoPRB2018Orbitalmagneticquadrupole, ShitadePRB2018Theoryorbitalmagnetic}
\begin{equation}
-e\frac{\partial M_{lk}}{\partial \mu} = I^{\rm{sea}}_{lk} + I_{lk}^{\rm{surf},1} + I_{lk}^{\rm{surf},2},
\end{equation}
where 
\begin{widetext}
\begin{align}
I^{\rm{sea}}_{lk} &= - \int \frac{d^2\bm{k}}{(2\pi)^2}\sum_{n\neq m} \frac{2\hbar e\text{Im}\left[\left\langle n,\bm{k}\right|v_{l}\left|m,\bm{k}\right\rangle \left\langle m,\bm{k}\right|m_{k}\left|n,\bm{k}\right\rangle \right]}{\left(\varepsilon_{n,\bm{k}}-\varepsilon_{m,\bm{k}}\right)^{2}}f_{n}, \\
I_{lk}^{\rm{surf},1} & = \int \frac{d^2\bm{k}}{(2\pi)^2}\sum_{n\neq m}\frac{2\hbar e}{3}\frac{\text{Im}\left[\left\langle n,\bm{k}\right|v_{l}\left|m,\bm{k}\right\rangle \left\langle m,\bm{k}\right|m_{k}\left|n,\bm{k}\right\rangle \right]}{\varepsilon_{n,\bm{k}}-\varepsilon_{m,\bm{k}}} \frac{\partial f_{n}}{\partial \varepsilon_{n,\bm{k}}},  \\
I_{lk}^{\rm{surf},2} & = -\int \frac{d^2\bm{k}}{(2\pi)^2} \sum_{n} \frac{\hbar e}{3}\epsilon_{ijk}\left(\frac{1}{4}\partial_{i}\Gamma_{n,lj}-v_{i,n}g_{n,lj}\right)\frac{\partial f_{n}}{\partial \varepsilon_{n,\bm{k}}}.
\end{align}
\end{widetext}
Here, $\bm{m}$ is the orbital magnetic dipole moment operator \cite{XiaoPRL2005BerryPhaseCorrection}, $\epsilon_{ijk}$ is the Levi-Civita symbol, $\Gamma_{n,lj}=\left\langle n,\bm{k}\right|\partial_{l}v_{j}\left|n,\bm{k}\right\rangle $ is the Hessian matrix element, $v_{i,n} = \langle n,\bm{k}|v_i |n,\bm{k}\rangle$, $f_n$ is the Fermi-Dirac distribution function for band $n$, and $g_{n,lj}=\text{Re}\sum_{m\left(\neq n\right)}A_{l,nm}A_{j,mn}$ is the quantum metric tensor with $A_{l,nm}$ the inter-band Berry connection. In zero temperature limit, $I_{lk}^{\rm{surf},2} = 0$ for the low-energy two-band Hamiltonian with rotational symmetry $C_n$ ($n=2,3,4,6$) \cite{FangPRL2012MultiWeylTopological} (Supplemental Material, Sec. S6 \cite{Supp}), leaving $I_{lk}^{\rm{surf},1}$ as the Fermi surface contribution in the main text. A flowing charge dipole $\bm{p}$ generates an orbital magnetic moment following $\bm{m}=\frac{1}{4}\left(\bm{p}\times\bm{v}-\bm{v}\times\bm{p}\right)$ \cite{FanNC2024IntrinsicdipoleHall}. For an inter-surface charge dipole, $\bm{m}=\frac{ed}{4}\hat{z}\times\left(\sigma_{z}\bm{v}+\bm{v}\sigma_{z}\right)=\frac{d}{2}\hat{z}\times\bm{j}^{D}$ \cite{FanNC2024IntrinsicdipoleHall}. Thus, 
\begin{equation}
I^{\rm{sea}}_{lk} = \epsilon_{kzj}\frac{d}{2}\sigma_{jl}^{D}.
\end{equation}
giving rise to the generalized Středa formula in insulators, i.e., Eq.~(\ref{eq:Streda_formula_1}), where $I_{lk}^{\text{surf,1}}=I_{lk}^{\text{surf,2}}=0$.

When $\mu$ lies at the valence band maximum, we calculate the integral of $I_{lk}^{\rm{surf},1}$ by the same parameterization method as mentioned in Appendix \hyperref[Appendix-B]{B}, and obtain
\begin{equation}
I_{lk}^{\rm{surf},1} \propto \frac{R_{0,0} \rm{sign}(\Delta_0)}{\big|A^2-2B\Delta_0 +2D |\Delta_0|\big|}.
\end{equation}
Here, $R_{0,0}$ is given by Eq.~(\ref{eq:B_R00}). If $R_{0,0} \neq 0$, $I_{lk}^{\rm{surf},1}$ is ill-defined at CTPT with $\Delta_0 \rightarrow 0$, yielding the failure of the generalized Středa formula Eq.~(\ref{eq:Streda_formula_1}) for a conventional gapless CTPT. On the contrary, in quasisymmetry-enriched CTPT with nontrivial charge, $R_{0,0}=0$ and thus $I_{lk}^{\rm{surf},1}=0$, enforced by quasisymmetry $\Pi$ [Eq.~(\ref{eq:quasisymmetry_BHZ})] in the BHZ model. This validates the generalized Středa formula for dipole Hall conductivity at the gapless criticality. See Supplemental Material, Sec. S6 \cite{Supp} for details.

\end{document}